\documentclass{aa}
\usepackage{graphicx}
\usepackage{txfonts}
\begin{document}
   \title{A search for solar-like oscillations in the Am star HD~209625\thanks{Based
   on observations collected at the 3.6-m telescope at La Silla Observatory (ESO, Chile: program 073.D-0527 and 075.D-0614)}}

   \author{F. Carrier
          \inst{1,2}
          \and
          P. Eggenberger
	  \inst{3,2}
	  \and
          J.-C. Leyder
	  \inst{3} \fnmsep \thanks{FNRS Research Fellow}
	  \and
          Y. Debernardi
	  \inst{2}
	  \and
	  F. Royer
	  \inst{4}
          }

   \offprints{F. Carrier}

   \institute{Instituut voor Sterrenkunde, Katholieke Universiteit Leuven, 200D Celestijnenlaan,
              B-3001 Leuven, Belgium\\
	      \email{fabien@ster.kuleuven.be}
         \and
             Observatoire de Gen\`eve, Universit\'e de Gen\`eve, 51 Chemin des Maillettes,
             CH-1290 Sauverny, Switzerland \\
	     \email{yves@deber.ch}
	 \and
	     Institut d'Astrophysique et de G\'eophysique, Universit\'e de Li\`ege, 17 All\'ee du 6 Ao\^ut, B\^at. B5c, B-4000 Li\`ege, Belgium\\
	     \email{eggenberger@astro.ulg.ac.be, leyder@astro.ulg.ac.be}
	 \and
	     Observatoire de Paris-Meudon, GEPI, 5 Place Jules Janssen, F-92195 Meudon cedex, France\\
	     \email{frederic.royer@obspm.fr}
             }

   \date{Received ; accepted }

 
  \abstract
   {}
   {The goal is to test the structure of hot metallic stars, and in particular the structure
   of a near-surface convection zone using asteroseismic measurements. Indeed, stellar models
   including a detailed treatement of the radiative diffusion predict the existence of a near-surface convection zone in order to correctly reproduce the anomalies in surface abundances that are observed in Am stars.}
   {The Am star \object{HD 209625} was observed with the \textsc{Harps} spectrograph 
   mounted on the 3.6-m telescope at the ESO La Silla Observatory (Chile) during 9 nights
   in August 2005. This observing run allowed us to collect 1243 radial velocity (RV) measurements\thanks{The 1243 radial velocities are only available in electronic form
   at the CDS via anonymous ftp to cdsarc.u-strasbg.fr (130.79.128.5)
or via http://cdsweb.u-strasbg.fr/cgi-bin/qcat?J/A+A/}, with a standard 
   deviation of 1.35\,m\,s$^{-1}$.}
   {The power spectrum associated with these RV measurements does not present any excess. 
Therefore, either the structure of the external layers of this star does not allow excitation of solar-like
oscillations, or the amplitudes of the oscillations remain below 20--30\,cm\,s$^{-1}$ (depending on their frequency range).
}
   {}


   \maketitle
%

\section{Introduction}

In the spectral range from A- to F-type stars, almost 70\,$\%$ of 
the non-chemically peculiar (CP) stars
are currently classified as $\delta$~Sct variables, while most non-variable stars are
Am-type stars. Although observations do not completly rule out variability in Am stars
(or A-type metallic-line stars), they do
rigorously constrain it: either variability is extremely rare, or the pulsations
are of very low amplitude.

At their arrival on the ZAMS, Am stars have a HeII convection zone, typical of A--F
stars. Due to their low rotational velocity, mixing processes are not efficient
enough to prevent He from settling down. Helium leaves the external layers 
and is therefore no longer able to excite pulsations
typical for the classical instability strip. 
Moreover, selective radiative acceleration 
causes the accumulation of iron-peak elements where these elements dominate the opacity, 
and leads to the appearance of iron-peak convection zones
centered at a temperature of approximately 200\,000\,K (Turcotte et al. \cite{trmc00}).  
When assuming that there is a sufficient overshoot to homogenize the surface regions
between the H, He and iron-peak convection zones, the predicted surface abundance  
variations closely resemble the abundance anomalies observed in Am stars
(Richer et al. \cite{rich00}). Detailed models of Am stars thus predict the existence
of a surface convection zone which may extend deeply into the
external layers of the star. For instance, in a typical A-type star of 1.7\,M$_\odot$,    
the surface convection zone is predicted to only include the outer $10^{-8}$ to $10^{-9}$\,M$_\odot$
of the star, while it may include the outer $10^{-6}$ to $10^{-5}$\,M$_\odot$
of the star for an Am star with iron-peak convection zones (Richard et al. \cite{rich01}). 
The total mass of the convection zone for such an Am star
is then found to be very similar to the mass of the surface convection zone of a 1.4\,M$_\odot$ star 
at solar metallicity which is able to drive solar-like oscillations as observed
in the subgiant Procyon A (see e.g. Marti\'c et al. \cite{mar04} and Eggenberger et al. \cite{ecbb04}).   

In order to probe the structure of the predicted convection zone due to iron overabundances, 
we performed high-accuracy asteroseismic observations of an Am star. Indeed, depending on its structure, this convection zone may be able to drive solar-like oscillations
similar to the ones detected for less massive stars 
(see e.g. the review by Bedding \& Kjeldsen \cite{bed06}).

Typical A-type stars show only a few spectral lines (as opposed to F or G stars)
and rotate fast, which prevents us from obtaining
high-precision radial velocity (RV) measurements. 
However, this is not the case for Am stars characterized
by under-abundances of calcium and scandium, and by
over-abundances of iron-group elements.
The large number of metallic absorption lines in Am spectra 
is particularly valuable to obtain precise radial velocities, needed to detect
solar-like oscillations.
Moreover, the rotational velocity distribution of Am stars is very different
from that of regular A stars (Debernardi, \cite{dd02}): their mean rotational velocity
lies below 50\,km\,s$^{-1}$, thus some Am stars have very low
rotational velocities, a necessary condition to obtain accurate radial velocities.
Therefore, Am stars are the only hot stars
for which the high radial velocity accuracy needed to detect solar-like oscillations can be reached. 

The target selected
is the Am SB1 binary \object{HD~209625} (A5/A9/F2; with a $V$ magnitude
of 5.28). This object is representative of the Am main sequence stars
while having a quite low rotational velocity $v \sin i=9.6$\,km\,s$^{-1}$ (deduced from our radial velocity measurements).

In this paper, we report on Doppler observations of \object{HD~209625} performed with the \textsc{Harps} spectrograph,
and resulting in the determination of a very high-accuracy time series for an Am star. 
 
The observations and the associated data reduction are presented in Sect.~\ref{odr}, along with
a study of the binarity of the target.
The power spectrum analysis is discussed in Sect.~\ref{psa}, followed by
the conclusions in Sect.~\ref{conclu}.

\section{Observations and study of binarity}
\label{odr}
The Am star \object{HD~209625} was observed during 4 half-nights in September 2004 and during 9 nights in August 2005 with the \textsc{Harps} 
spectrograph (Pepe et al. \cite{pmr02}; see e.g. Carrier \& Eggenberger \cite{ce06} for solar-like oscillations with this instrument) mounted on the 3.6-m
telescope at the La Silla
Observatory (ESO, Chile). We took consecutive exposures, each lasting between 140 and 300\,s (depending on the airmass and the extinction), and separated by dead times of 30\,s.
In total, 430 and 1243 spectra were collected (for the first and second run respectively), with a typical signal-to-noise ratio (S/N) ranging from 100 to 300 at 530\,nm.
In parallel to the stellar exposures, the spectrum of a thorium lamp carried by a second fiber was simultaneously recorded,
in order to monitor the spectrograph's stability.
The spectra obtained were extracted on-line. The radial velocities
were computed online by weighted cross-correlation with a numerical mask constructed from a G2 dwarf spectrum. In addition, they were also
determined by weighted cross-correlation with different masks built from more appropriate synthetic spectra (spectral type between A5 and F2, see Carrier et al. \cite{cardeb}) and by the optimum-weight procedure (Connes \cite{c85}, Carrier et al. \cite{ceb05}), although without significant gain.
The use of a G2 template can introduce a small systematic shift in the radial velocities but without consequence for the variability analysis of this star. 

   \begin{figure}
   \resizebox{\hsize}{!}{\includegraphics{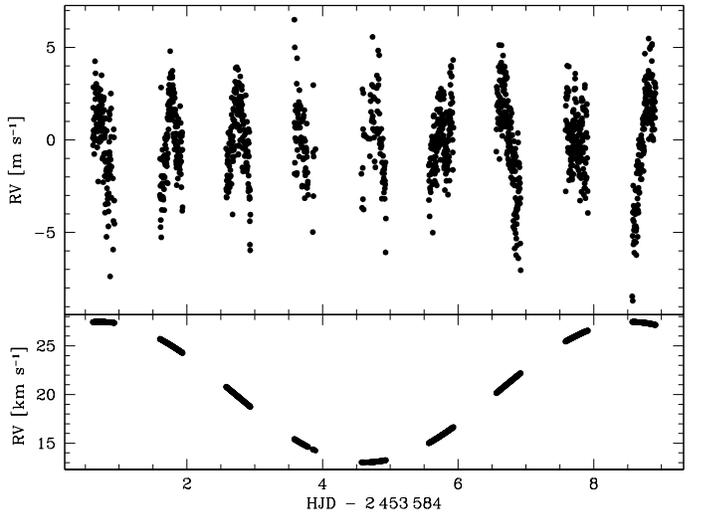}}
   \caption{\textsc{Harps} radial velocity measurements of \object{HD~209625}. 
The bottom panel shows the radial velocities due to the orbital motion, with the parameters of Table~\ref{orb}.
The top panel presents the RV measurements from which the sinusoidal fit was removed,
and after the average of each night was set to zero.}
              \label{figrv}%
    \end{figure}

The Am star \object{HD~209625} being a SB1 binary, it is necessary to correct the observations
for the Keplerian movement before studying the presence of p-mode oscillations. 
In order to accurately determine the period of this binary system, we used all \textsc{Harps} radial velocity measurements,
along with previous \textsc{Coravel} measurements (Debernardi \cite{dd02}) and with data from Abt and Levy (\cite{al85}). Then, the other parameters were determined using only the \textsc{Harps} measurements,
because they are far more accurate. 
The eccentricity was found to be very small (0.001) and was thus fixed to zero.
The derived orbital parameters are listed in Table~\ref{orb}.
The RV modulation due to the binarity of the system is shown in the bottom panel of
Fig.~\ref{figrv}, while the top panel shows the RV measurements during the second run
(the 9 nights in August 2005) with the orbital
modulation removed. 
  
\begin{table}
\caption{Orbital elements derived for \object{HD~209625} when taking into account \textsc{Harps}, \textsc{Coravel} and Abt and Levy measurements (see text for more details).}
\begin{center}
\begin{tabular}{lc}
\hline
\hline
Parameter  &  Value \\ \hline
$P$ (days)    & 7.83238 $\pm$ 0.00002\\ 
$T_{0}$ (HJD)& 53420.2304 $\pm$ 0.0001 \\ 
$e$            & 0 (fixed)\\ 
$\omega$ (deg)& 0 (fixed)\\ 
$V_0$ (km s$^{-1}$)& 20.2630 $\pm$ 0.3 \\ 
$K_1$ (km s$^{-1}$) & 7.2150 $\pm$ 0.4\\
$a_{1} \sin i$ (Gm)  &0.77708 $\pm$ 0.00004\\ 
$f_1(m)$ (10$^{-3}$ M$_{\odot}$)        & 0.30550 $\pm$ 0.00005\\ 
$\sigma_{(O-C)}$ (m s$^{-1}$) & 2.1 \\
\hline
\label{orb}
\end{tabular}
\end{center}
\end{table}

We note that most nights seem to exhibit the same shape: an increase of the RV, followed by a decrease.
In fact, the radial velocities are correlated both with the airmass and with the ratio of the flux in the blue 
part of the spectrum to the flux in the red part.
This modulation is related to the atmospheric extinction.  At the accuracy of the m\,s$^{-1}$, all spectrum orders
show systematic differences in velocity. The time series computed from the cross-correlations, which are 
weighted means by the flux of correlations of each order,
are thus correlated to the flux change.
No corrections were applied in the online 
pipeline to take into account such atmospheric extinction changes. However, this effect should be corrected in the new online version of the pipeline (F. Pepe, private communication).

Although part of the analysis is based on both observing runs, the second run measurements are mainly used, because
the complicated window function of the first run consisting of 4 half-nights leads to a less precise result, however
in full agreement with the first run. 

A very small variation, related to either intrinsic variations of the star or perhaps to a tiny eccentricity,
remains in the data (i.e. shift of 1--2 m\,s$^{-1}$ between nights). For this reason,
the mean value for each night was also subtracted. This could slightly modify the power spectrum at very low frequencies, but has no influence in the range where oscillations are expected
(Carrier et al. \cite{cedw05}).

\begin{table}
\caption{Distribution of the 1243 Doppler measurements during the second run.}
\begin{center}
\begin{tabular}{clll}
\hline
\hline
Date & Nb of spectra & Duration & Dispersion $\sigma$  \\ 
 &  & (hours) & (m\,s$^{-1}$) \\ \hline
2005/08/01 & 134 & 7.63  & 2.09 \\
2005/08/02 & 154 & 8.01  & 1.99 \\
2005/08/03 & 169 & 8.57  & 1.94 \\
2005/08/04 & 71  & 7.46  & 1.97 \\
2005/08/05 & 72  & 8.66  & 2.27 \\
2005/08/06 & 170 & 8.77  & 1.60 \\
2005/08/07 & 167 & 8.61  & 2.67 \\
2005/08/08 & 156 & 7.90  & 1.63 \\
2005/08/09 & 150 & 8.23  & 3.05 \\ \hline
\label{tabrv}
\end{tabular}
\end{center}
\end{table}

\section{Power spectrum analysis}
\label{psa}
 \begin{figure*}
   \resizebox{\hsize}{!}{\includegraphics{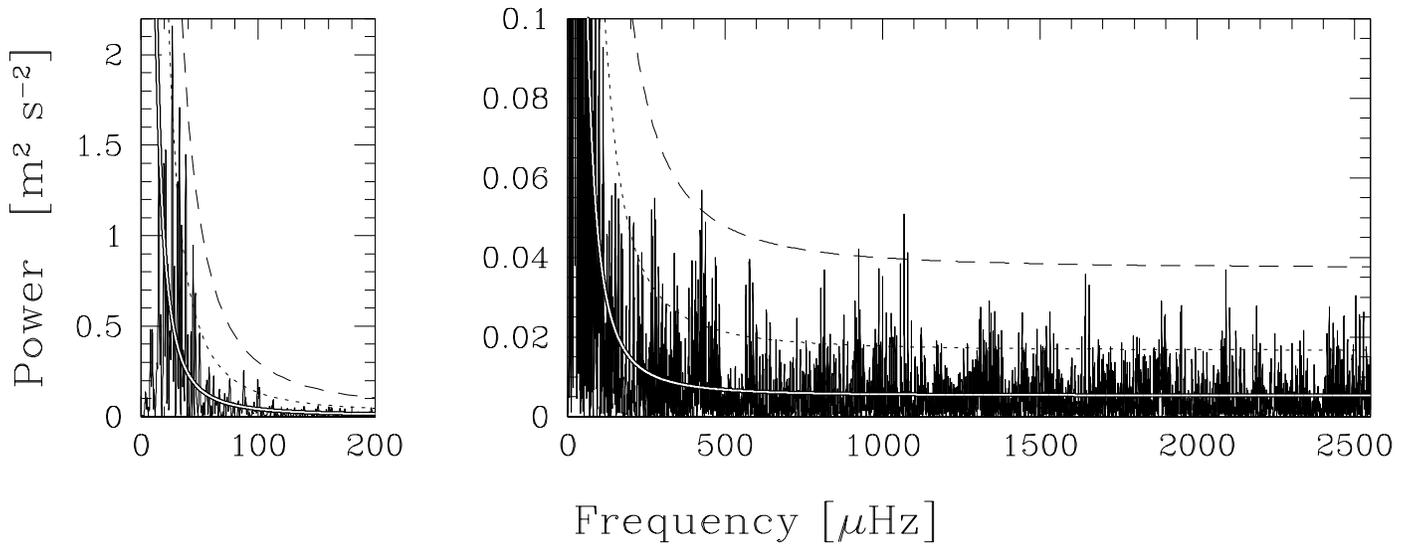}}
   \caption{Power spectrum of \object{HD~209625} (right), with a zoom on the low-frequency part (left). The white line indicates the mean noise level, while dotted and dashed lines
   represent an amplitude of 2\,$\sigma$ and 3\,$\sigma$ respectively.}
              \label{figtf}%
    \end{figure*}
The power spectrum of the velocity time series was computed using
the Lomb-Scargle modified algorithm (Lomb \cite{lomb}; Scargle \cite{scargle}),
and is shown in Fig.~\ref{figtf}. Its time scale gives a formal resolution
of 1.39\,$\mu$Hz.
As typically observed in such a power spectrum, the noise is made of two distinct components:
\begin{itemize}
\item At high frequencies the noise is flat, indicative
of the Poisson statistics of photon noise. The mean white noise level $\sigma_{\mathrm{pow}}$ calculated between 1.5 and 2.5~mHz 
is 0.00511\,m$^2$\,s$^{-2}$, corresponding to
$\sigma_{\mathrm{amp}} = \sqrt{\sigma_{\mathrm{pow}} * \pi / 4}=6.3$\,cm\,s$^{-1}$ in amplitude (Kjeldsen \& Bedding \cite{kb}). With 1243 measurements, this high frequency noise 
corresponds to $\sigma_{RV}\,=\,\sqrt{N \sigma_{\mathrm{pow}} /4 }\,=\,1.3$~m\,s$^{-1}$. This radial velocity uncertainty is exceptionally low for an A-star, but remains nevertheless
over the photon noise uncertainty limit which is estimated by the pipeline at 0.5--1.\,m\,s$^{-1}$. This small discrepancy could either
be explained by an intrinsic variability of the star (such as granulation or stellar activity) or by an instrumental effect.
\item Towards the lowest frequencies, the power spectrum scales inversely with the square of the frequency, 
as expected for instrumental instabilities. 
\end{itemize}
The mean noise is indicated by the white line in Fig.~\ref{figtf}, along with the 2\,$\sigma$ value
(dotted line) and 3\,$\sigma$ value (dashed line).

\begin{figure}
   \resizebox{\hsize}{!}{\includegraphics{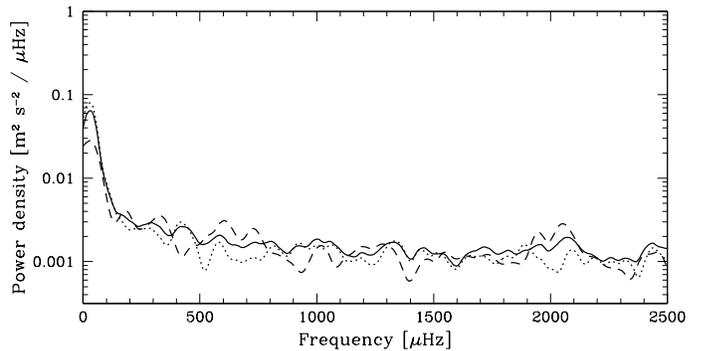}}
   \caption{Power density of \object{HD~209625} (with the power axis in logarithmic scale). The dashed, dotted and full lines correspond respectively to the first, the second
   and both runs.}
              \label{figtfdens}%
    \end{figure}

As discussed by Kjeldsen \& Bedding (\cite{kb}), the conversion to the power density requires multiplying the power spectrum
by the effective observing time, calculated by integrating under the spectral window. 
Density spectra were built for both runs separately and also
for the combined data. The effective observing time has a value of 0.87, 2.50 and 3.56\,d
for the first run, the second one and the combined data, respectively. The smoothed power density spectra are displayed in Fig.~\ref{figtfdens}. 
The density spectrum of the first run is the most affected by noise because of the smaller number of measurements.

Scaling from the solar case (Kjeldsen \& Bedding \cite{kb}), the frequency of the greatest mode of \object{HD~209625} would be expected to lie near
500--600\,$\mu$Hz, although this value cannot be trusted as the internal structure of an Am star is totally
different from the one of a solar--type star.
Solar-like oscillations are detected neither in the power spectrum (see Fig.~\ref{figtf}) 
nor in a smoothed version of the density spectrum (see Fig.~\ref{figtfdens}) at frequencies
below 2500\,$\mu$Hz.
Moreover, Fig.~\ref{figtf} shows that only two peaks (near 400 and 1100\,$\mu$Hz)
have an amplitude greater than 3\,$\sigma$, a result in agreement with pure noise (Carrier
et al.~\cite{ceb05}).
     
Although no oscillations could be detected by searching for a power excess directly in the periodogram or in the power density spectrum, we also tried to identify a comb-like structure in the power spectrum
corresponding to solar-like oscillations, and in particular to $\ell$\,=\,0 modes, by computing autocorrelations (region by region and with different thresholds).
However, no autocorrelation showed the presence of such a comb-like structure.
   \begin{figure}
   \resizebox{\hsize}{!}{\includegraphics{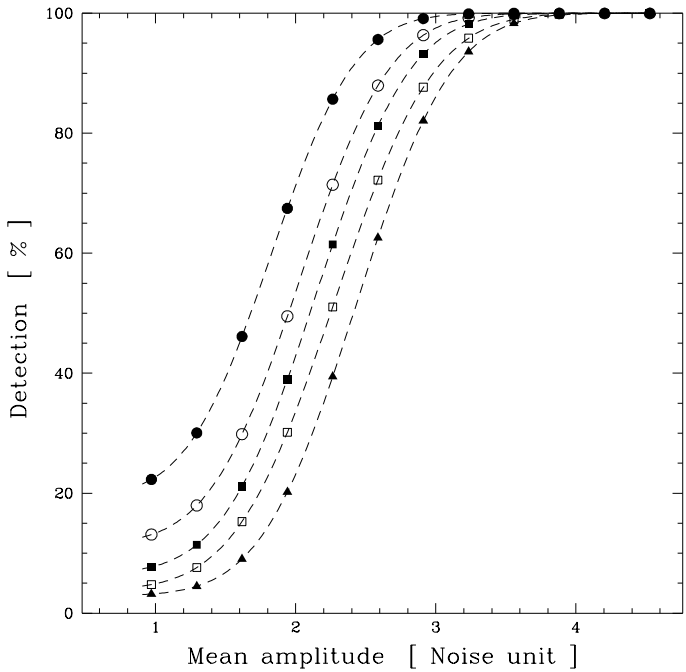}}
   \caption{Simulations giving the probabiltiy of detection of the oscillations versus the mean amplitude of the three highest modes. The oscillations
   are detected if an excess appears above a certain threshold in the density spectrum. This threshold varies between 3\,$\sigma$ (black circle)
   and 5\,$\sigma$ (black triangle), in steps of 0.5\,$\sigma$ (see text for more details on the simulations).}
              \label{figsimu}%
    \end{figure}

In order to determine a tighter upper limit on the oscillations, 
we also performed simulations: artificial modes were introduced while reproducing the overall noise level.
The frequencies of these injected modes were separated by 20\,$\mu$Hz, while their amplitudes were described by a Gaussian enveloppe with a $\sigma$ of 50\,$\mu$Hz; this corresponded to a few peaks over a small frequency range.
The main free parameter in these simulations was the amplitude of the Gaussian.
The other two parameters were also modified to check the independance of the results: no significant change was observed provided no more than 3 or 4 introduced peaks had a non-negligible amplitude.
A Gaussian noise, randomly drawn (using Monte-Carlo simulations) and adjusted to give the same value as in the observed power spectrum, was then added to the time series.
Using the same time sampling as observed, radial velocities were determined by assuming an infinite lifetime for the modes. 
Since the power spectrum of solar-like oscillations is known to depend on the poorly-known lifetime of the modes, we thus worked with the density spectrum, which is far less sensitive to the lifetime value.
The density spectrum is expected to be flat if there is no signal. The noise introduces some modulation, characterized by the $\sigma_{\mathrm{rms}}$ of the density spectrum.
For each value of the introduced Gaussian amplitude, 1\,000 different simulations were performed, leading to 1\,000 density spectra. Different thresholds were used to determine the detection probability, ranging from $3\sigma_{\mathrm{rms}}$ to $5\sigma_{\mathrm{rms}}$ by steps of $0.5\sigma_{\mathrm{rms}}$. Then, the number of density spectra where peaks larger than a given threshold were observed provided the detection probability, for this given threshold and Gaussian amplitude of the modes. Finally, the mean of the 3 highest amplitude modes was computed. 
Fig.~\ref{figsimu} shows the detection probability of modes with a given mean amplitude, for different detection thresholds.
In the observed density spectrum (Fig.~\ref{figtfdens}), no peak above $3.3\sigma_{rms}$ is seen.
This $3.3\sigma_{rms}$ threshold, when compared to the results of the simulations, means that oscillations with a mean amplitude of 3.05 (respectively 2.70) times the noise can be detected with a probability of 99\,\% (respectively 95\,\%).

\section{Conclusion}
\label{conclu}
We do not detect any solar-like oscillations in the Am star \object{HD~209625}: either this object does not exhibit any oscillations, or
the amplitude of its oscillations is below our detection level. 

To unambiguously detect a power excess due to solar-like oscillations (at the 99\,\% confidence level),
the noise in the Fourier transform must be, at least, 3.05 times smaller than the amplitude of the oscillations.
Therefore, the oscillation amplitude is below 20\,cm\,s$^{-1}$ if the star oscillates 
at frequencies above 800\,$\mu$Hz, and could reach 
30\,cm\,s$^{-1}$ for oscillations with frequencies near 200\,$\mu$Hz. 
Such solar-like oscillations therefore remain, if they exist, very limited in amplitude.
 
This non-detection of solar-like oscillations with a clear upper limit puts strong constraints on the structure of near-surface convection zones predicted by
theoretical models of Am stars. Indeed, these observations allow us to rule out all models of the Am star \object{HD~209625} that predict a surface convection zone that would be able
to excite solar-like oscillations with an amplitude larger than 30\,cm\,s$^{-1}$.

\begin{acknowledgements}
      Part of this work was supported by the Swiss National Science Foundation.
JCL acknowledges support through the XMM-INTEGRAL PRODEX project and IAP contract P5/36. 
\end{acknowledgements}


\begin{thebibliography}{}

\bibitem[1985]{al85} Abt, H.A., \& Levy, S.G. 1985, ApJS, 59, 229
\bibitem[2006]{bed06} Bedding, T.R., \& Kjeldsen, H. 2006, Memorie della Societa Astronomica Italiana, 77, 384
\bibitem[2002]{cardeb} Carrier, F., Burki, G., \& Burnet, M., 2002, A\&A, 385, 488
\bibitem[2005a]{ceb05} Carrier, F., Eggenberger, P., \& Bouchy, F. 2005a, A\&A, 434, 1085
\bibitem[2005b]{cedw05} Carrier, F., Eggenberger, P., D'Alessandro, A., \& Weber, L. 2005b, New Astron., 10, 315
\bibitem[2006]{ce06} Carrier, F., \& Eggenberger, P. 2006, A\&A, 450, 695
\bibitem[1985]{c85} Connes, P. 1985, Ap\&SS, 110, 211
\bibitem[2002]{dd02} Debernardi, Y. 2002, Ph.D. Thesis, Geneva Observatory, No 3362, "Binarity of Am stars"
\bibitem[2004]{ecbb04} Eggenberger, P., Carrier, F., Bouchy, F., \& Blecha, A. 2004, A\&A, 422, 247
\bibitem[1995]{kb} Kjeldsen, H., \& Bedding, T.R. 1995, A\&A, 293, 87
\bibitem[1976]{lomb} Lomb, N.R. 1976, Ap\&SS, 39, 447
\bibitem[2004]{mar04} Marti\'c, M., Lebrun, J.-C., Appourchaux, T., \& Korzennik, S.G. 2004, A\&A, 418, 295 
\bibitem[2000]{rich00} Richer, J., Michaud, G., \& Turcotte, S. 2000, ApJ, 529, 338 
\bibitem[2001]{rich01} Richard, O., Michaud, G., \& Richer, J. 2001, ApJ, 558, 377 
\bibitem[1982]{scargle} Scargle, J.D. 1982, ApJ, 263, 835
\bibitem[2002]{pmr02} Pepe, F., Mayor, M., Rupprecht, G., et al. 2002, Messenger, 110, 9
\bibitem[2000]{trmc00} Turcotte, S, Richer, J., Michaud, G., \& Christensen-Dalsgaard, J 2000, A\&A, 360, 603

\end{thebibliography}
\end{document}